\documentclass{iau} 
\usepackage{graphicx}
\usepackage[export]{adjustbox}

\newcommand{\Msun}{\mbox{\,$M_{\odot}$}}

\title[Properties of Andromeda's Stellar Halo] 
{Recent Results from SPLASH: Chemical Abundances and Kinematics of Andromeda's Stellar Halo}

\author[Gilbert, Beaton, and Dorman]   
{Karoline M. Gilbert$^1$,
Rachael Beaton$^2$,
Claire Dorman$^3$,
\and the SPLASH collaboration}

\affiliation{$^1$Space Telescope Science Institute, \\ 3700 San Martin Dr.,
Baltimore, MD 21218, USA \\ email: {\tt kgilbert@stsci.edu} \\[\affilskip]
$^2$The Observatories of the Carnegie Institution of Washington \\ 813 Santa Barbara St., Pasadena, CA 91101, USA \\email: {\tt rbeaton@obs.carnegiescience.edu} \\[\affilskip]
$^3$UCO/Lick Observatory, University of California at Santa Cruz, \\1156 High Street, Santa Cruz, CA 95064, USA \\ email: {\tt cdorman@ucolick.org}
}

\pubyear{2015}
\volume{317}  
\jname{To be published in: The General Assembly of Galaxy Halos: Structure, Origin and Evolution}
\editors{A. Bragaglia, M. Arnaboldi, M. Rejkuba, \& D. Romano}
\begin{document}

\maketitle

\begin{abstract}
Large scale surveys of Andromeda's resolved stellar populations have revolutionized our view of this galaxy over the past decade. The combination of large-scale, contiguous photometric surveys and pointed spectroscopic surveys has been particularly powerful for discovering substructure and disentangling the structural components of Andromeda. The SPLASH (Spectroscopic and Photometric Landscape of Andromeda's Stellar Halo) survey consists of broad- and narrow-band imaging and spectroscopy of red giant branch stars in lines of sight ranging in distance from 2 kpc to more than 200 kpc from Andromeda's center.  The SPLASH data reveal a power-law surface brightness  profile extending to at least two-thirds of Andromeda's virial radius (\cite[Gilbert et al.\ 2012]{gillbert2012}), a metallicity gradient extending to at least 100 kpc from Andromeda's center (\cite[Gilbert et al.\ 2014]{gilbert2014}), and evidence of a significant population of heated disk stars in Andromeda's inner halo (\cite[Dorman et al.\ 2013]{dorman2013}).  We are also using the velocity distribution of halo stars to measure the tangential motion of Andromeda (Beaton et al., in prep).

\end{abstract}

\firstsection 
\section{Introduction}

The Milky Way (MW) and Andromeda (M31) provide the two best opportunities for in depth studies of stellar halos. Moreover, they provide complementary perspectives. While our internal vantage point provides us with exquisite detail of the inner halo of the MW, it is difficult to study the outer halo due to the large distance uncertainties, low stellar densities and vast survey areas required for even a partial view of the halo.  Conversely, our external vantage point provides a global view of the halo of Andromeda. While Andromeda's distance has historically precluded measurements that are commonplace for the MW’s halo, the last decade has seen tremendous progress.  This has largely been due to the photometric and spectroscopic observations obtained by the SPLASH (Spectroscopic and Photometric Landscape of Andromeda's Stellar Halo) and PAndAS (Pan-Andromeda Archaeological Survey; \cite[McConnachie et al.\ 2009]{mcconnachie2009}) collaborations.  

The SPLASH collaboration has utilized the Mosaic camera on the  Kitt Peak 4\,m Mayall Telescope to obtain broad-band ($M$ and $T_2$) and narrow-band (DDO51) imaging and the DEIMOS spectrograph on the Keck~II 10\,m telescope to obtain spectroscopy of individual M31 stars.  The surface-gravity sensitive DDO51 imaging enables the selection of stars that are likely to be red giants at the distance of Andromeda, greatly increasing the efficiency of the spectroscopic observations.  The SPLASH collaboration has imaged 78 fields and obtained $>$\,20,000 stellar spectra in Andromeda's disk, dwarf galaxies, and halo, in fields ranging from 2\,--\,230~kpc in projected distance from Andromeda's center.

This dataset has led to the discovery and characterization of Andromeda's extended, metal-poor stellar halo (\cite[Guhathakurta et al.\ 2005]{guhathakurta2005}, \cite[Gilbert et al.\ 2006]{gilbert2006}, \cite[Kalirai et al.\ 2006a]{kalirai2006halo}, \& \cite[Courteau et al.\ 2011]{courteau2011}), and has been used to measure the global properties of the halo (\cite[Gilbert et al.\ 2012 \& 2014]{gilbert2012,gilbert2014}) and to characterize the inner stellar halo and disk (\cite[Dorman et al.\ 2012, 2013, \& 2015]{dorman2012, dorman2013, dorman2015}).  It has been used to study the properties of Andromeda's dwarf satellites (\cite[Majewski et al.\ 2007]{majewski2007},  \cite[Kalirai et al.\ 2009]{kalirai2009}, \cite[Howley et al.\ 2008]{howley2008}, \cite[Kalirai et al.\ 2010]{kalirai2010}, \cite[Tollerud et al.\ 2012]{tollerud2012}, \cite[Ho et al.\ 2012]{ho2012} \& \cite[Howley et al.\ 2013]{howley2013}), to identify and characterize tidal debris features (\cite[Guhathakurta et al.\ 2006]{guhathakurta2006}, \cite[Kalirai et al.\ 2006b]{kalirai2006gss}, \cite[Gilbert et al.\ 2007, 2009a, 2009b]{gilbert2007,2009a,2009b}), and led to the discovery of the continuation of Andromeda's giant southern stream (\cite[Gilbert et al.\ 2007]{gilbert2007}, \cite[Fardal et al.\ 2008, 2012]{fardal2008,2012}).

\begin{figure}[bt]
\begin{center}
 \includegraphics[width=3.2in]{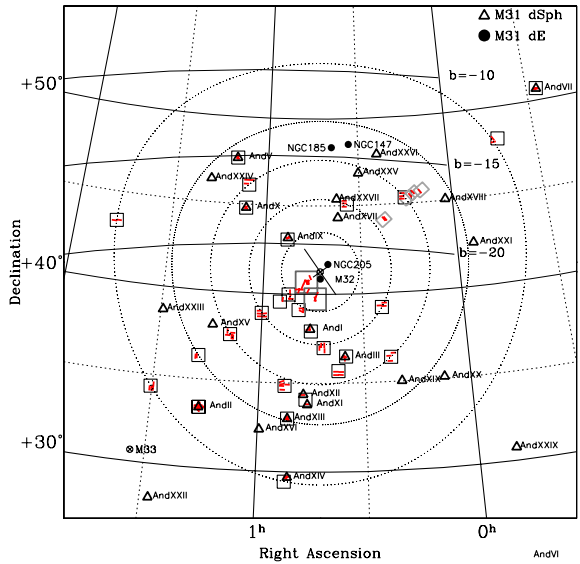} 
 \caption{Location of the SPLASH photometric (larger rectangles) and spectroscopic (small, narrow rectangles) observations used in the measurement of the surface brightness and metallicity profiles of Andromeda's stellar halo.  Figure from \cite[Gilbert et al.\ (2012)]{gilbert2012}.}
   \label{fig1}
\end{center}
\end{figure}

\section{Surface Brightness Profile and Metallicity Gradient}
SPLASH spectroscopic and photometric observations in 38 halo fields (Fig.\,\ref{fig1}) have been combined to measure the radial surface brightness and metallicity profile of Andromeda's stellar halo.  These lines-of-sight span all quadrants of the halo and range from 9\,--\,230~kpc in projected distance from Andromeda's center.  The stellar spectra were used to identify secure samples of Andromeda red giant branch stars (removing Milky Way dwarf star contaminants; \cite[Gilbert et al.\ 2006]{gilbert2006}) and to identify stars associated with kinematically cold tidal debris features (\cite[Gilbert et al.\ 2007, 2009b, \& 2012]{gilbert2007, gilbert2009b, gilbert2012}).   

\begin{figure}[t!]
\begin{center}
 \includegraphics[width=2.2in]{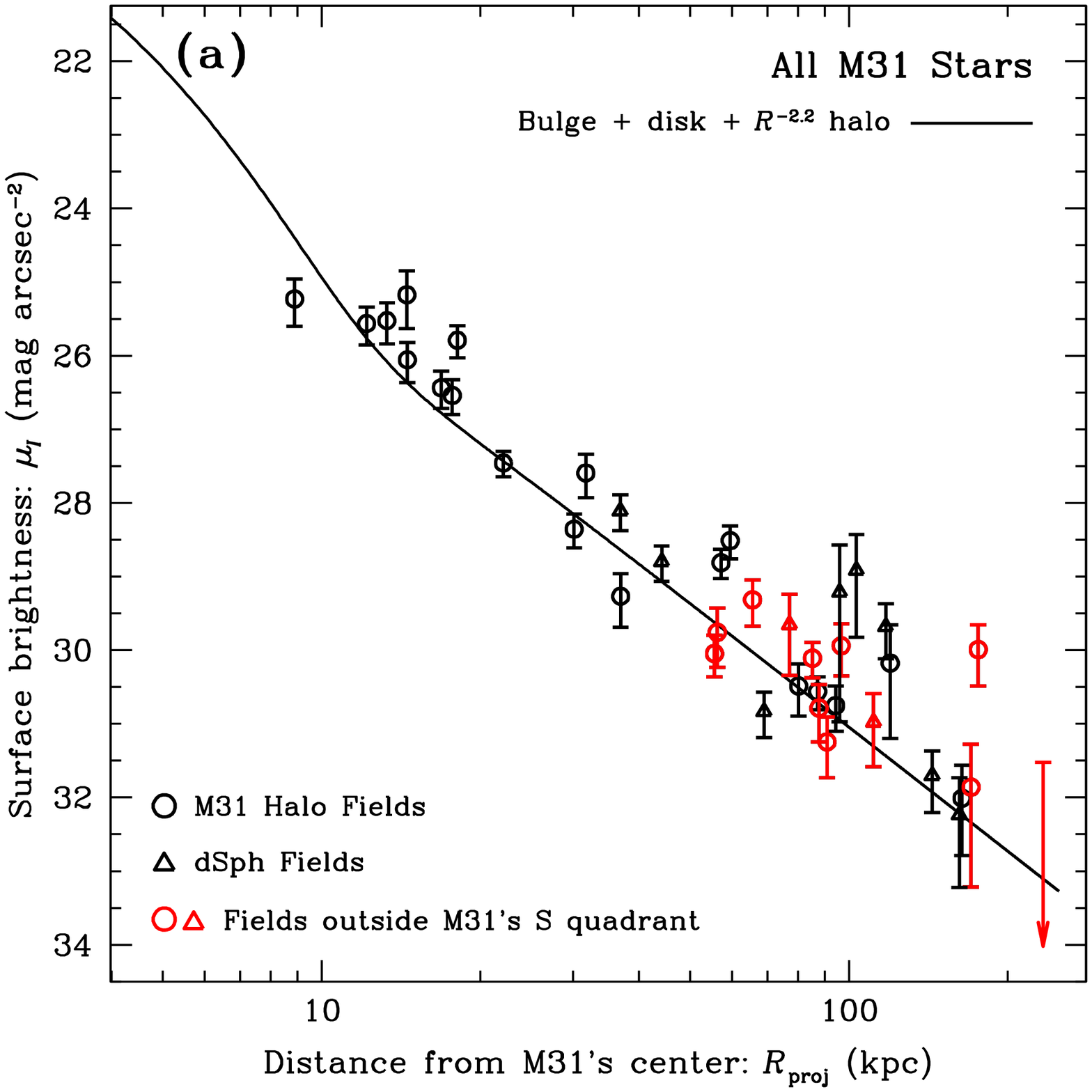} 
 \includegraphics[width=2.2in]{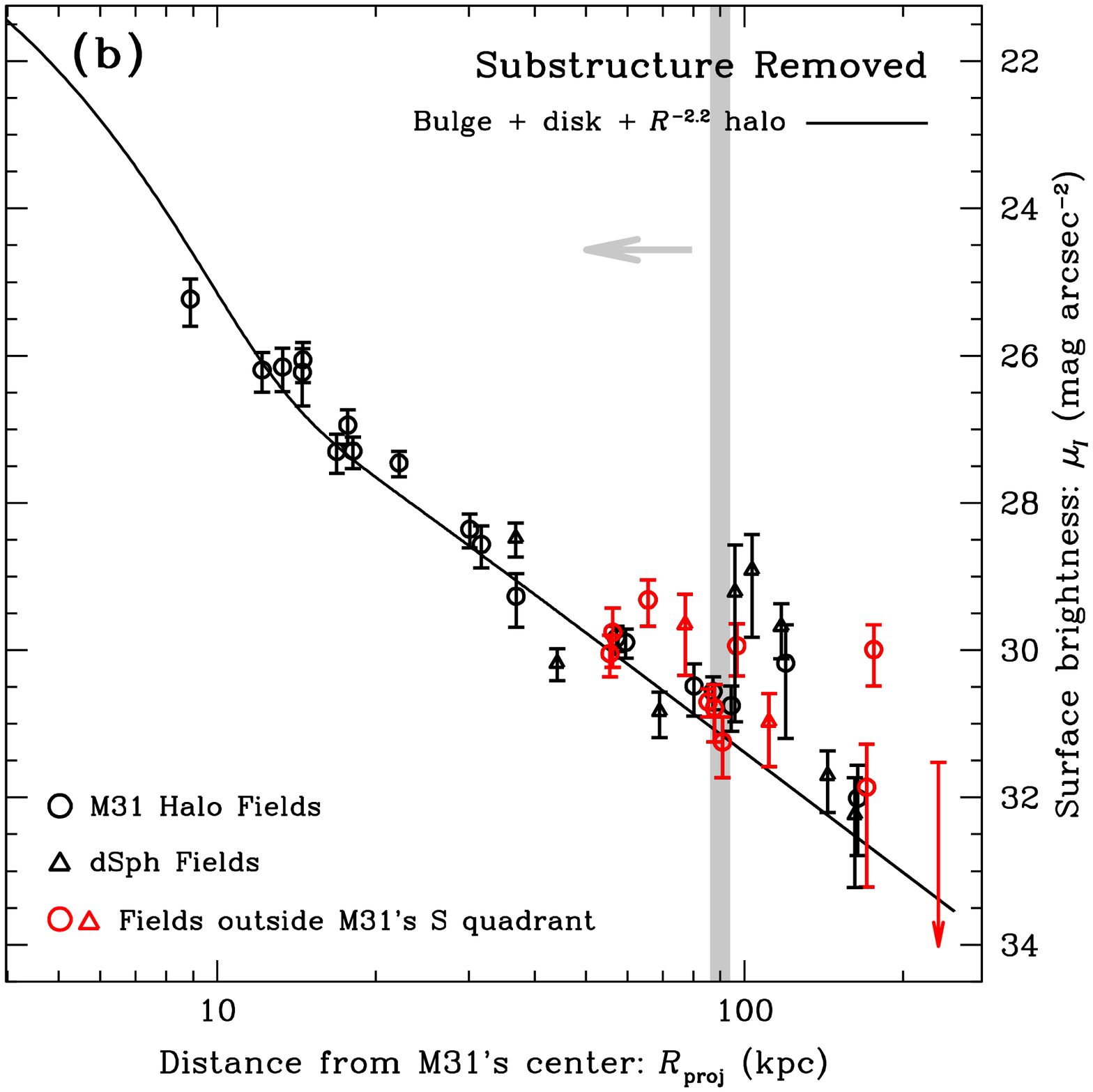} 
 \caption{Surface brightness profile of Andromeda's stellar halo, with (left) and without (right) stars associated with kinematically cold tidal debris features.  Andromeda's stellar halo is consistent with a single power-law with an index of $-2.2$ over a radial range of 10\,--\,175~kpc in projected distance from Andromeda's center. Tidal debris features have only been identified in fields less than 90~kpc from Andromeda's center: the low number of Andromeda stars in the outer halo fields ($R_{\rm proj} > 90$~kpc) prevent identification of multiple kinematical components.  Figures are from \cite[Gilbert et al.\ (2012)]{gilbert2012}.}
   \label{fig2}
\end{center}
\end{figure}

\begin{figure}[t!]
\begin{center}
 \includegraphics[width=2.2in]{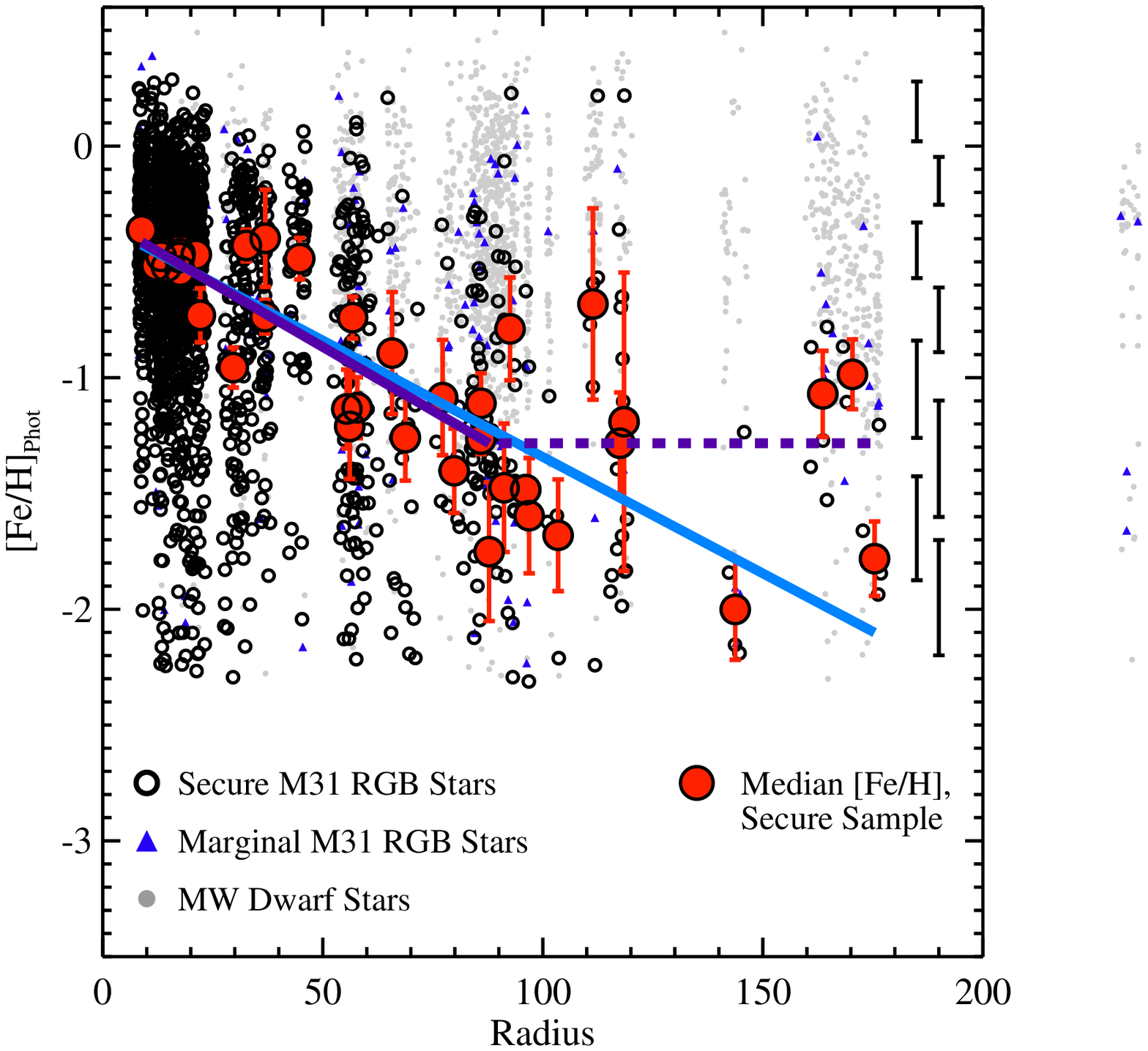} 
 \includegraphics[width=2.2in]{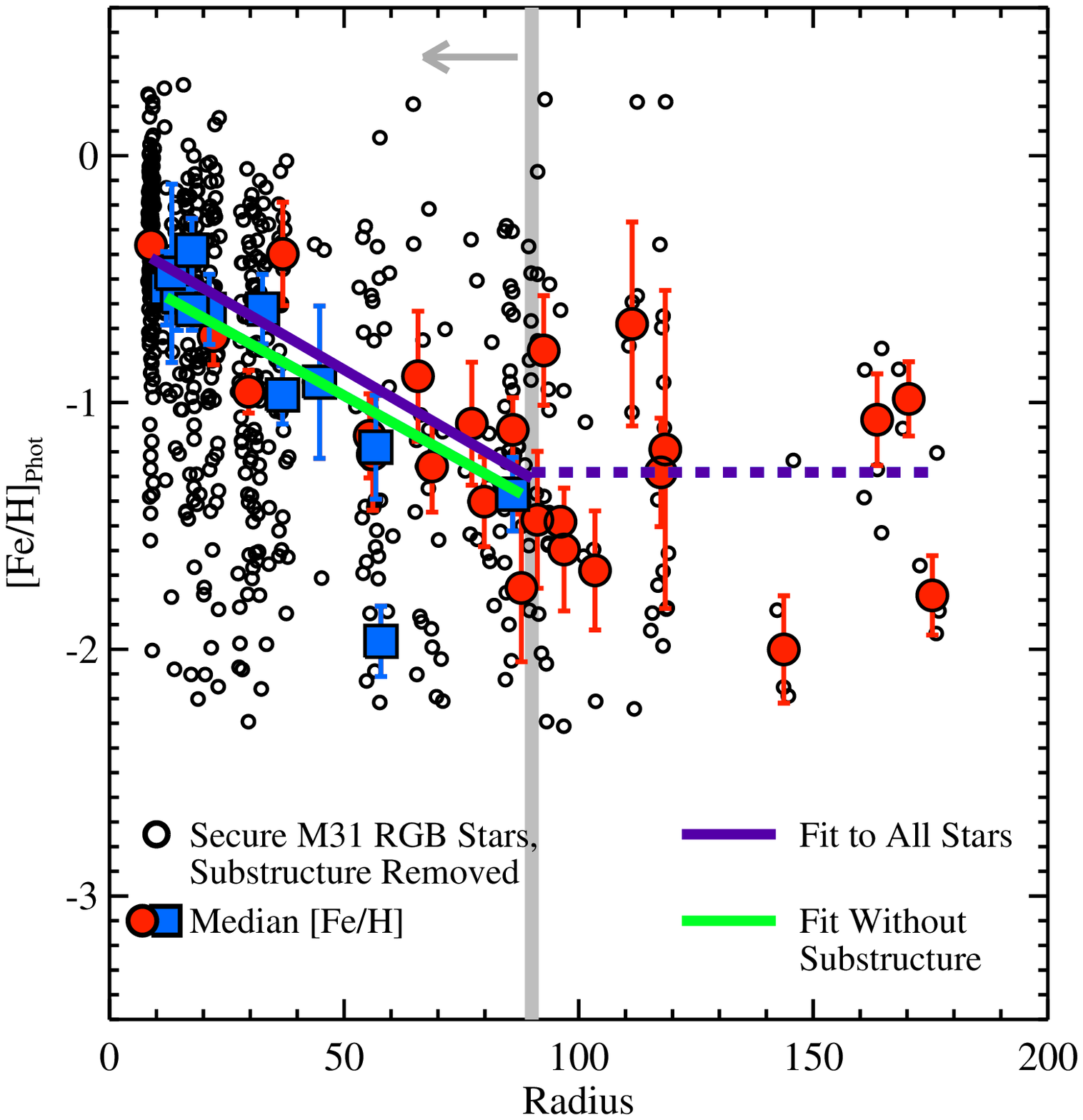} 
 \caption{Metallicity profile of Andromeda's stellar halo, with (left) and without (right) stars associated with kinematically cold tidal debris features.  As in Fig.\,\ref{fig2}, tidal debris features have only been identified and removed in fields less than 90~kpc from Andromeda's center.  In the right panel, large squares denote the median [Fe/H] in the fields where tidal debris features have been identified and removed.   Figures are from \cite[Gilbert et al.\ 2014]{gilbert2014}.}
   \label{fig3}
\end{center}
\end{figure}

The surface brightness profile of Andromeda's stellar halo is consistent with a single power-law extending to a projected distance of more than 175~kpc from Andromeda's center ($\sim 2/3$ the virial radius;  Fig.\,\ref{fig2}).  The surface brightness profile is derived from the ratio of counts of spectroscopically confirmed Andromeda red giant branch stars to Milky Way dwarf stars (\cite[Gilbert et al.\ 2012]{gilbert2012}).  The right panel shows the surface brightness profile once tidal debris features have been removed, using the results of maximum-likelihood, multi-Gaussian fits to each field's velocity distribution.  Tidal debris features have been identified in half of the SPLASH spectroscopic fields within 90~kpc of Andromeda's center.  

Andromeda's stellar halo shows clear evidence of a metallicity gradient, extending to $\sim$100~kpc (Fig.\,\ref{fig3}), with a total decrease of $\sim$1~dex (\cite[Gilbert et al.\ 2014]{gilbert2014}).  The [Fe/H] estimates are based on a comparison of the star's position in the color-magnitude diagram with stellar isochrones (assuming an age of 10~Gyr and the solar value of [$\alpha$/Fe]).  In the right panel, stars that may be associated with tidal debris features have been removed, and a gradient of $-1$~dex over 100~kpc in radius remains.  This shows that the metallicity gradient in Andromeda's stellar halo is {\it not} driven by the tidal debris features included in the SPLASH dataset.  
For a subset of spectra with the highest S/N, spectroscopic estimates of [Fe/H] based on the equivalent width
 of the Calcium Triplet were compared with the photometric estimates of [Fe/H].   On average the spectroscopic and photometric estimates agree, and the same gradient in [Fe/H] with radius is found.

Broad inferences about Andromeda's merger history can be made by comparing the surface brightness and metallicity profiles to simulations of stellar halo formation.  The lack of a downward break in the surface brightness profile may indicate that Andromeda has undergone a fairly large number of recent, low-mass accretion events (\cite[Gilbert et al.\ 2012]{gilbert2012}).  The large-scale metallicity gradient extending over 100~kpc may indicate that the majority of the stars in the halo were contributed by one to a few early, relatively massive ($\sim 10^9$\Msun) accretion events (see discussion in \cite[Gilbert et al.\ 2014]{gilbert2014}).  However, to make more concrete statements about the luminosity function and time of accretion of the satellites that formed Andromeda's stellar halo, we will need to measure the $\alpha$-element abundances of halo stars.  \cite[Vargas et al.\ (2014a,b)]{vargas2014a,vargas2014b} published the first [$\alpha$/Fe] measurements of Andromeda stars (including 4 halo stars and measurements in 9 dwarf galaxies).  The SPLASH dataset provides a rich archive for measuring [$\alpha$/Fe] in many more halo fields.

\begin{figure}[bt]
\begin{center}
 \includegraphics[width=2.6in]{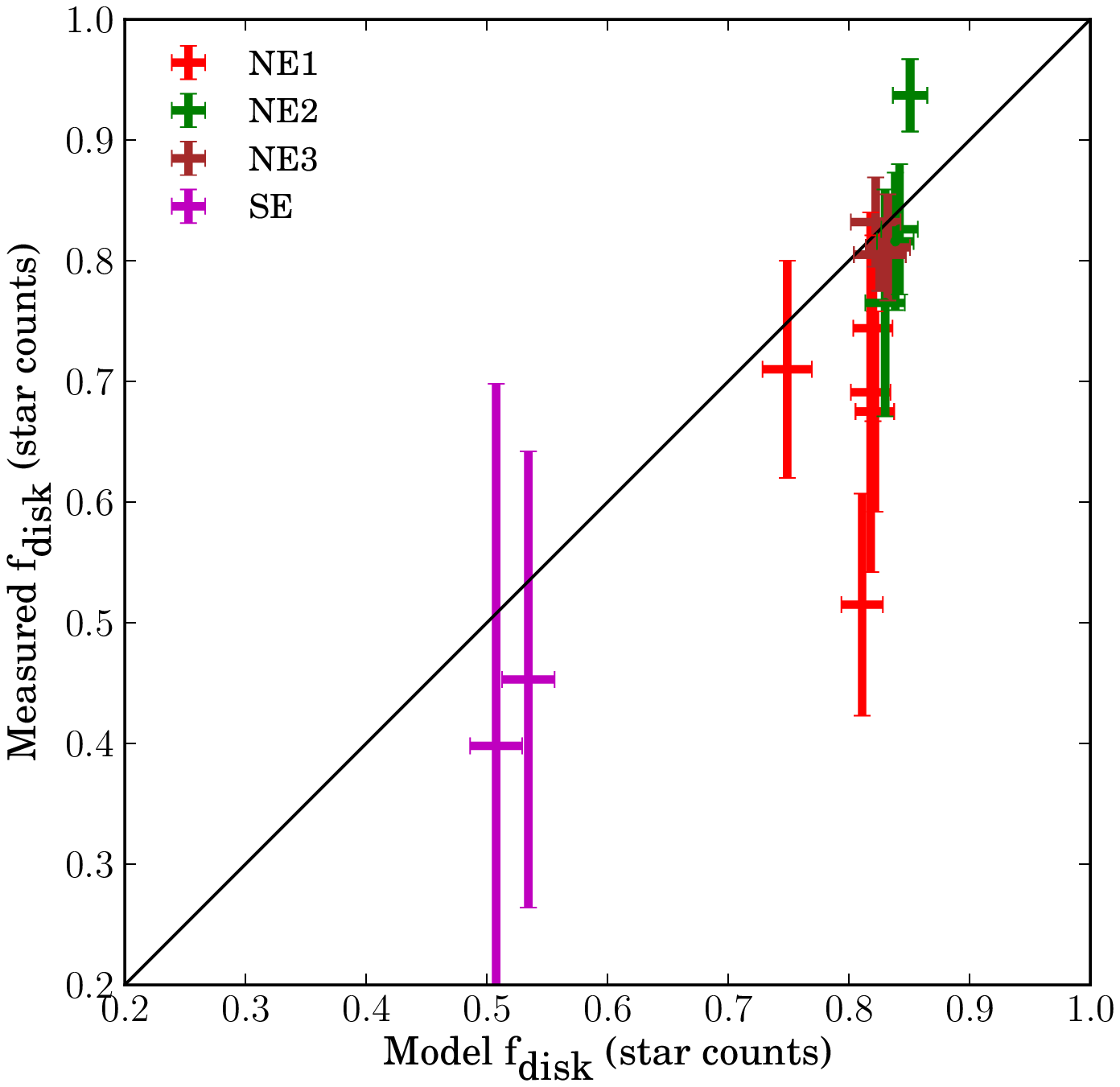} 
 \includegraphics[width=2.6in]{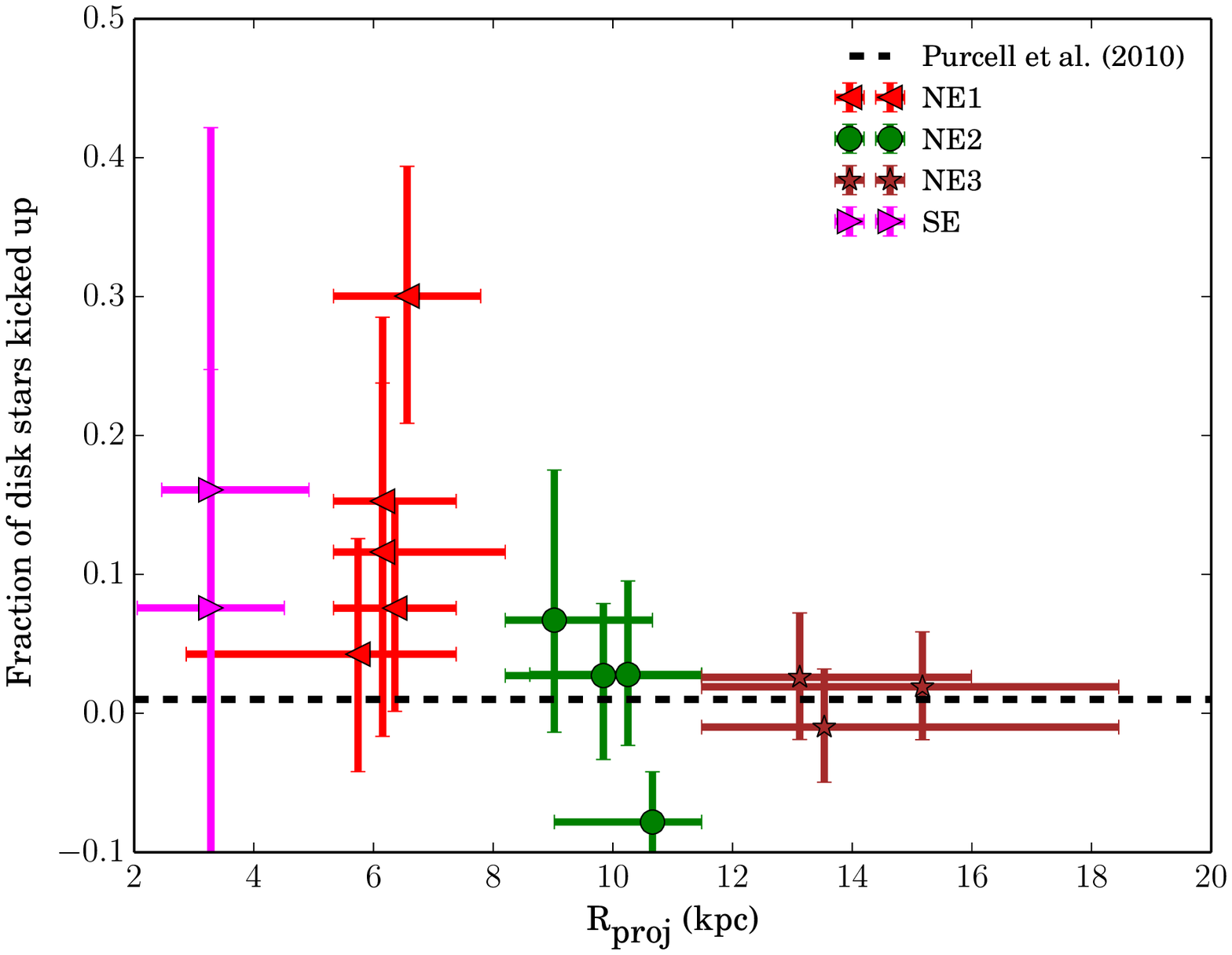} 
 \caption{Evidence for a kicked-up disk component in Andromeda's inner halo.  {\it Left:} Comparison of the disk fraction from a kinematical analysis of the velocity distribution of stars (measured $f_{\rm disk}$) with the disk fraction favored by a model that includes unresolved surface brightness data and the stellar luminosity function, as well as the stellar velocity distribution (model $f_{\rm disk}$).  The disk fraction favored by the model is systematically higher than the measured dynamically cold disk fraction, suggesting there is a population of dynamically hot (``kicked-up'') disk stars.  {\it Right:}  In order to reconcile the observed stellar luminosity function and the stellar kinematics, a fraction of the stars with a disk-like luminosity function must be dynamically hot.  This panel shows the dynamically hot fraction of disk stars as a function of radius.  Vertical error bars denote $1\sigma$ uncertainties while horizontal error bars show the full radial range of each spatial region shown in the figure.  Figures are from \cite[Dorman et al.\ (2013)]{dorman2013}.}
   \label{fig4}
\end{center}
\end{figure}

\section{Kinematics of M31's Inner Spheroid}
\cite[Dorman et al.\ (2013)]{dorman2013} combined stellar kinematics from SPLASH ($>5000$ spectra), the stellar luminosity function derived from resolved HST imaging (from the PHAT survey; \cite[Dalcanton et al.\ 2012]{dalcanton2012}), and unresolved surface photometry in Andromeda's disk to model the relative strength of each of the structural components (bulge, disk, and halo) as a function of radius.  The best-fitting models favor a disk fraction that is systematically higher than the disk fraction measured solely from an analysis of the stellar velocity dispersion (Fig.\,\ref{fig4}).  The number of stars with a disk-like luminosity function is 5.2\%$\pm 2.1$\% larger than the number of stars with dynamically cold, disk-like kinematics.  This implies Andromeda's inner regions contain a population of stars with a disk-like luminosity function but spheroid-like kinematics.  This is the first direct evidence that there are stars that were born in Andromeda's disk and subsequently dynamically heated into the halo; they are now kinematically indistinguishable from Andromeda's inner stellar halo.  The number of stars that have been dynamically heated are consistent with that expected from simulations (see discussion in \cite[Dorman et al.\ 2013]{dorman2013}).

\section{Tangential Motion of M31 From Halo Stars}

\begin{figure}[bt]
\begin{center}
 \includegraphics[width=2.4in]{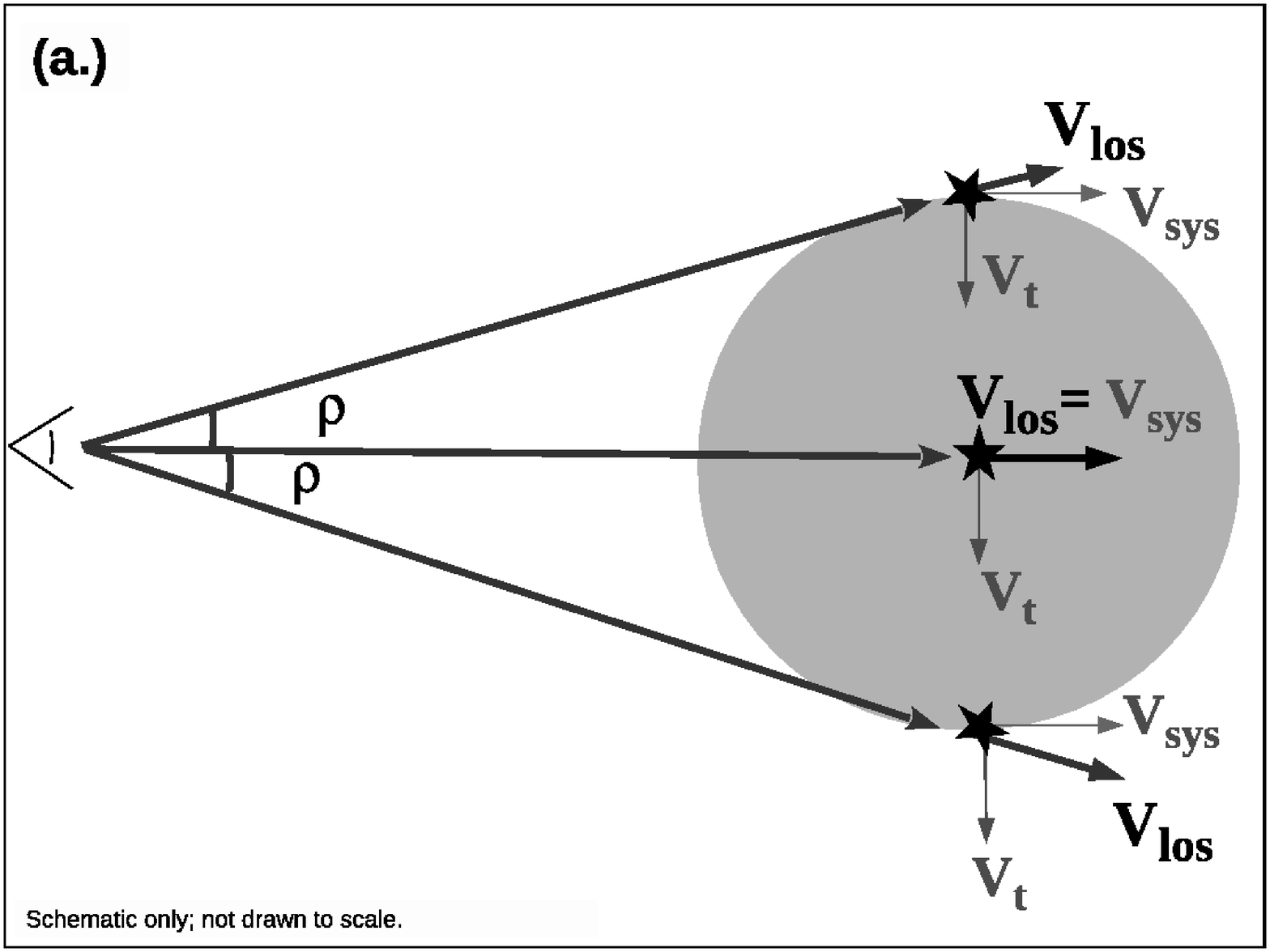} 
 \includegraphics[width=2.2in]{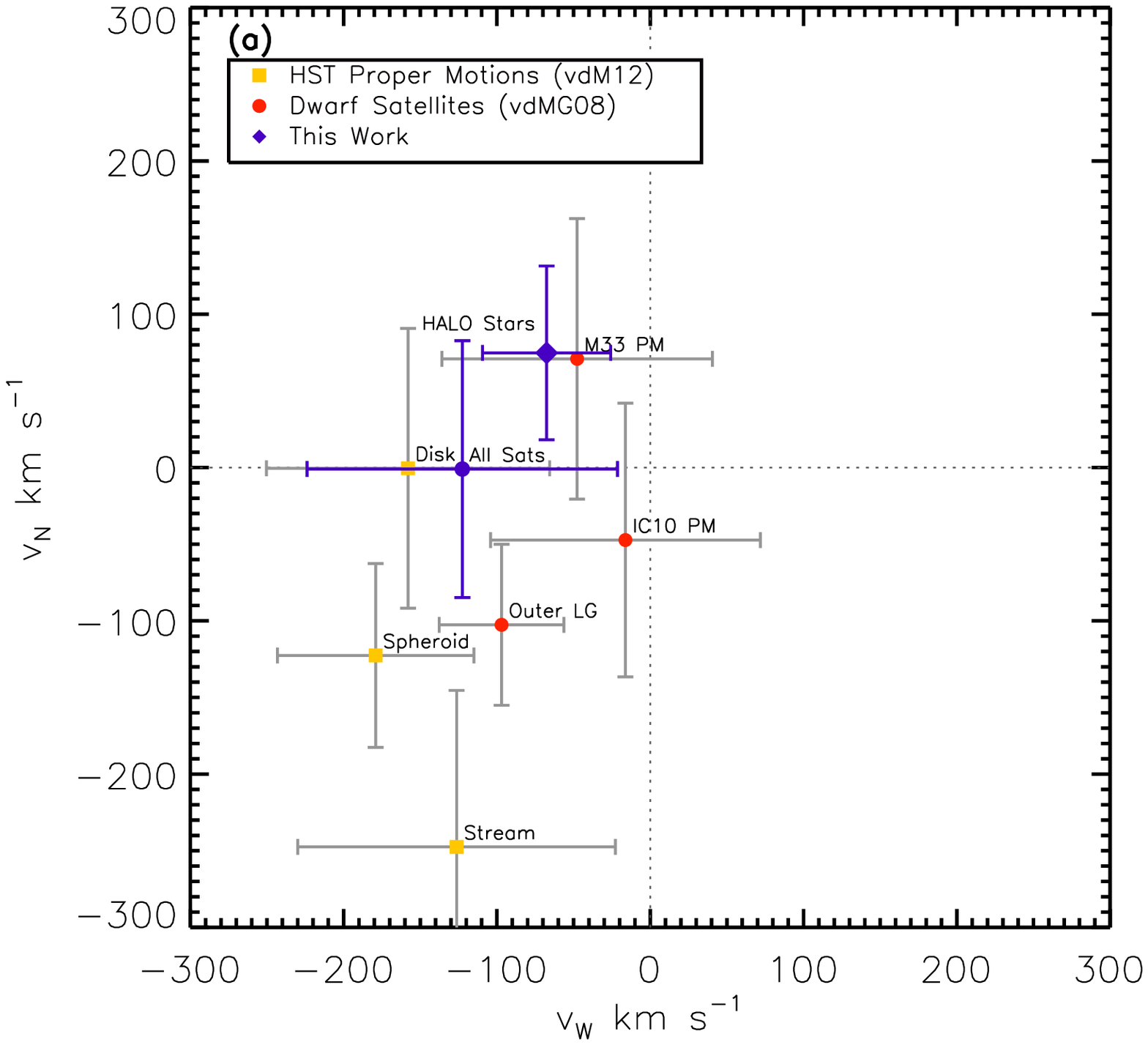} 
 \caption{\textit{Left:} Schematic illustration of the center-of-mass motion of an object in the sky plane along two dimensions ---  $v_{\rm sys}$ along the line-of-sight to the center of mass and $v_{\rm t}$ perpendicular to $v_{\rm sys}$ ---  and the effect of observing tracers at lines-of-sight at widely separated angles, $\rho$. 
For a tracer at $\rho\sim0$, the line-of-sight of the center of mass and of the tracer are aligned, whereas for $\rho >$\,0, this is no longer the case (i.e., $v_{los} \neq v_{\rm sys}$) and the magnitude of the difference depends on both the radial ($\rho$) and azimuthal ($\phi$) location of the tracer.  \textit{Right:} Measurements of Andromeda's tangential motion;  $v_{W}$ and $v_{N}$ 
are the plane of sky motions to the West and North.  Measurements from different sources are compared: direct HST proper motions (square) and statistical tests using the line-of-sight velocities of dwarf satellite galaxies (circles) and halo stars (diamond). In the on-line proceedings, each point is color-coded based on the source of the measurement: \cite[van der Marel et al.\ (2012; yellow)]{vandermarel2012}, \cite[van der Marel \& Guhathakurta (2008; red)]{vandermarel2008}, or Beaton et al. (in prep; blue).}
   \label{fig5}
\end{center}
\end{figure}

Knowledge of the orbit of Andromeda is central to understanding 
the past, current, and future dynamical state of the Local Group.  
This requires determining Andromeda's three dimensional motion,   
an extremely challenging measurement. Recently, \cite[Sohn et al.\ (2012)]{sohn2012} 
used HST imaging obtained over a long time baseline to 
measure Andromeda's tangential motion via the observed proper motion of 
Andromeda's stars.   

Indepedent confirmation of Andromeda's transverse velocity 
can be made using the kinematics of halo populations.  
This technique relies on the projection of the three dimensional 
center-of-mass motion on the line-of-sight motions 
of bound objects (Fig.\,\ref{fig5}, \textit{left}). 
This technique is most powerful when applied over the large 
spatial baseline provided by tracer populations in Andromeda's halo: 
dwarf satellites,  globular clusters
(\cite[van der Marel \& Guhathakurta 2008]{vanderMarel2008}, 
\cite[van der Marel et al.\ 2012]{vanderMarel2012}) and halo stars. 

Beaton et al.\ (in prep) are employing 
the SPLASH dataset to make the first 
measurement of Andromeda's tangential motion using line-of-sight velocities of individual halo stars.
We also incorporate the kinematics of Andromeda's dwarf satellites 
and halo globular clusters. 
Our model's only assumption is that the tracer stars come from a single 
virialized component (\cite[van der Marel \& Guhathakurta 2008]{vanderMarel2008}). 
Following \cite[Gilbert et al.\ (2012)]{gilbert2012}, we exclude stars likely to 
belong to dwarf satellites or tidal debris features.  The greatest potential source of bias is from unidentified 
substructure.  However, the magnitude of this effect can be estimated 
empirically from the data and is added to the total uncertainty.  
We find broad agreement between our measurement of Andromeda's tangential motion and previous measurements (Fig.\,\ref{fig5}, \textit{right}). The error-weighted mean of all independent measurements is ($v_{W}$, $v_{N}$) = (-96.4 $\pm$ 22.4 km s$^{-1}$, -44.7 $\pm$ 25.4). 

\section{Summary}
The SPLASH collaboration has obtained a rich dataset of photometric and spectroscopic observations spanning the Andromeda system.  Among the many results from the SPLASH survey, four have been highlighted here.  This dataset has been used to measure the surface brightness and metallicity profile to two-thirds of Andromeda's virial radius (\cite[Gilbert et al.\ 2012 \& 2014]{gilbert2012,gilbert2014}).  In conjunction with the PHAT HST imaging of Andromeda's disk, it has been used to show that there is a significant population of kicked up disk stars in Andromeda's inner stellar halo (\cite[Dorman et al.\ 2013]{dorman2013}). Finally, it is providing an independent measurement of Andromeda's tangential motion (Beaton et al., in prep).

\end{document}